\documentclass[superscriptaddress,nofootinbib, amsmath,amssymb,preprintnumbers,prdfloatfix,twocolumn,PRL]{revtex4-2}
\usepackage{CJK}
\usepackage{graphicx}% Include figure files
\usepackage{dcolumn}% Align table columns on decimal point
\usepackage{upgreek}
\usepackage{amsmath}
\usepackage{bm}% bold math
\usepackage[colorlinks=true,pdfstartview=FitV,breaklinks=true]{hyperref}
\usepackage[dvipsnames,table]{xcolor}
\hypersetup{urlcolor=BlueViolet,
	    citecolor=Plum,
	    linkcolor=PineGreen}
\usepackage{lipsum}	    
\usepackage{titlesec}
\usepackage{etoolbox}% http://ctan.org/pkg/etoolbox
\usepackage[normalem]{ulem}
\usepackage{tabularx}
\usepackage{amssymb}
\usepackage{float}	
\usepackage{caption}
\usepackage{subcaption}
\usepackage{comment}
\usepackage[page]{appendix}
\usepackage{afterpage}
\usepackage{enumitem}
\usepackage{array}
\usepackage{listings}

\newcolumntype{P}[1]{>{\centering\arraybackslash}p{#1}}

\newcommand{\figref}[1]{Fig.~\ref{#1}}
\newcommand{\figsref}[2]{Figs.~\ref{#1}~and~\ref{#2}}

\newcommand{\secref}[1]{Sec.~\ref{#1}}
\newcommand{\appref}[1]{App.~\ref{#1}}
\newcommand{\tabref}[1]{Tab.~\ref{#1}}
\newcommand{\refref}[1]{Ref.~\cite{#1}}

\begin{document}

\title{Incentivizing supplemental math assignments and using AI-generated hints is associated with improved exam performance}

\author{Yifan Lu}
\email{yifanlu@ucla.edu}
\affiliation{Department of Physics and Astronomy, University of California Los Angeles,\\ Los Angeles, California, USA}

\author{K. Supriya}
\email{ksupriya@ceils.ucla.edu}
\affiliation{Center for Education, Innovation, and Learning in the Sciences, University of California Los Angeles,\\ Los Angeles, California, USA}

\author{Shanna Shaked}
\affiliation{Center for Education, Innovation, and Learning in the Sciences, University of California Los Angeles,\\ Los Angeles, California, USA}
\affiliation{Institute of Environment and Sustainability, University of California Los Angeles,\\ Los Angeles, California, USA}

\author{Elizabeth H. Simmons}
\affiliation{Department of Physics and Astronomy, University of California San Diego,\\ La Jolla, California, USA}

\author{Alexander Kusenko} 
\affiliation{Department of Physics and Astronomy, University of California Los Angeles,\\ Los Angeles, California, USA}
\affiliation{Kavli Institute for the Physics and Mathematics of the Universe (WPI), UTIAS \\The University of Tokyo, Kashiwa, Chiba, Japan} 

\begin{abstract}

Inequities in student access to trigonometry and calculus are often associated with racial and socioeconomic privilege, and often influence introductory physics course performance. To mitigate these disparities in student preparedness, we developed a two-pronged intervention consisting of (1) incentivized supplemental math assignments and (2) AI-generated learning support tools in the form of optional hints embedded in the physics homework assignments. Both interventions are grounded in the Situated Expectancy-Value Theory of Achievement Motivation, which posits that students are more likely to complete a task that they expect to do well in and whose outcomes they think are valuable. For the supplemental math assignments, the extra credit was scaled to make it worth more points for students with lower exam scores, thereby creating even greater value for students who might benefit most from the assignments. AI-generated hints were integrated into the homework assignments, thereby reducing or eliminating the cost to the student, in terms of time, energy, and social barriers or fear of judgment. Our findings indicate that both these interventions are associated with increased exam scores; in particular, the scaled extra credit reduced disparities in completion of supplemental math assignments. These interventions, which are relatively simple for any instructor to implement, are therefore very promising for creating more equitable undergraduate quantitative courses.

\end{abstract}

\maketitle

%%%%%%%%%%%%%%%%%%%%%%%%%%%%%%%%%%%%%%%%%%%%%%%%%%%%%%%%%%%%%%%%%%%%%%%%%%%%%%

\section{Introduction}
Students’ mathematical skills play an important role in shaping their learning and performance in undergraduate Introductory Physics courses \cite{hudson1977correlation,sadler2001success,meltzer2002relationship, PhysRevPhysEducRes.13.020137}. However, students enter these courses with wide variation in their prior exposure to advanced mathematics (i.e. trigonometry and calculus). Moreover, student access to advanced math courses is associated with racial and socioeconomic privilege, so that Black, Hispanic/Latine, and Native American students, and students from low-income backgrounds are less likely to have had the opportunity to take advanced math courses in high school \cite{battey_access_2013,bressoud_strange_2021,ackins_equity_2022, briones2023racialized}.

These disparities in learning advanced mathematics were further exacerbated during the COVID-19 pandemic when students spent an extended time learning remotely \cite{golden_what_2023, schultz_impact_2024}. Many high school students, especially Hispanic/Latine, Black, and Native American students and those from low-income backgrounds, experienced challenges that made learning more difficult. Some of the most commonly reported challenges included lack of quiet space to study or attend class, access to a computer and poor internet connection \cite{haderlein_disparities_2021, golden_what_2023, francom_technologies_2021, ong2020covid, vogels202053}. As the cohort of students who were in high school during the COVID-19 pandemic enter college, the negative impacts of remote learning on mathematical skills persist and must be addressed to help them succeed in college. 

One strategy to address the inequity in prior access to advanced mathematics courses is to provide supplemental instruction to students. This has been found to help students with less math preparation successfully complete Introductory Physics courses \cite{meling_supplemental_2013, desilva_enhancing_2018, PhysRevPhysEducRes.13.020137}. Most supplemental instruction takes the form of an optional in-person course or series of workshops focused on students’ problem solving skills in small groups \cite{burkholder_mixed_2021, desilva_enhancing_2018, meling_supplemental_2013}. This can be quite time-consuming and resource-intensive. Using online learning tools offers an alternative less resource-intensive way to approach supplemental instruction but relies more heavily on student motivation and might require students to pay additional fees \cite{PhysRevPhysEducRes.13.020137, PhysRevPhysEducRes.13.010122}. For example, Forrest et al. \cite{PhysRevPhysEducRes.13.020137} offered an optional asynchronous online math tutorial to Introductory Physics students who scored less than 65\% on a Math diagnostic exam and found that students who completed the online tutorial were four times more likely to pass the course. However, this tutorial was not free and only about half of the students completed the online tutorial, even though completing the tutorial was associated with a small amount of course points \cite{PhysRevPhysEducRes.13.020137}. Thus, motivating students to participate in optional online supplemental math instruction can be challenging \cite{devore_challenge_2017, PhysRevPhysEducRes.13.010122}.

Extra credit points and course credit points are often used to motivate students to complete supplemental math material. Course credit points especially can encourage a larger proportion of students to engage with the optional support offered \cite{PhysRevPhysEducRes.13.010122}. However, one potential drawback of offering credit for optional assignments is that it might disproportionately benefit students that are already doing well in the course and might not benefit as much from completing them \cite{PhysRevPhysEducRes.13.010122, hardy2002extra, harrison2011students}. For example, Mikula \& Heckler \cite{PhysRevPhysEducRes.13.010122} found that when offered extra credit for completing optional math assignments in an introductory physics course, 67\% of students with a high grade in the course completed the assignments compared to only 40\% of students with a low grade. More concerningly, credit associated with optional assignments might disproportionately benefit students with higher socioeconomic status and fewer responsibilities, who are less likely to need to work full time during college \cite{supriya2024optional}. Scaling the extra credit points so that students with lower scores in the course would benefit more from the optional supplemental assignments offers one way to mitigate this drawback.

An alternative to separate optional supplemental instruction is incorporation of hints within the usual course assignments. Intelligent tutoring systems that students could use for homework have been in place for almost two decades now. These systems have been shown to be effective in improving student learning at the K-12 as well as college level for many disciplines such as mathematics \cite{singh_feedback_2011, feng_implementing_2023, sale_learning_2019, cody_impact_2022}, accounting \cite{johnson_intelligent_2009}, and programming \cite{phothilimthana_high-coverage_2017}.  However, the development of these systems was time-intensive and required programming expertise. The emergence of Large-language Models (LLMs) such as GPT-4 over the past couple of years has made this much easier \cite{rus2024large}. For example, AI-generated hints are offered in all the online homework assignments on kudu.com. The hints are based on the text of the chapter the students are studying. The effectiveness of such hints generated by AI in supporting student learning are relatively unknown, but it holds much promise.

In this study, we compared the effectiveness of (1) optional assignments containing supplemental math material with scaled extra credit and (2) AI-generated hints on regular course assignments in supporting student learning and reducing inequities in student performance in an Introductory Physics course at a public university in the southwestern United States.

\subsection{Theoretical Framework: Expectancy-Value Theory of Achievement Motivation}
Our interventions in this study are derived from the Situated Expectancy-Value Theory of Achievement Motivation \cite{eccles_expectancy-value_2020, eccles_expectancies_1983}. This theory (earlier called Expectancy-Value Theory of Achievement Motivation) was initially developed in Ref. \cite{atkinson1957motivational} to understand risk-taking behavior. It was adapted for education by Jacquelynne Eccles and colleagues who were studying achievement behavior shaping sex differences in math achievement among students in fifth through twelfth grades \cite{eccles_expectancies_1983}. Over the past few decades, this theory has been widely applied across many educational contexts, including undergraduate education in STEM \cite{perez2014role, robinson2019motivation, Wu02012020, appianing2018development}. For example, Perez et al. \cite{perez2019science} applied expectancy-value theory to understand how students' beliefs in their science competence, task values, and perceived costs may coexist, exploring which combinations may be most relevant for STEM persistence and achievement.
According to this theory, students' motivation to engage in an activity is shaped by their expectations of success as well as by the perceived value of the activity \cite{eccles_expectancies_1983, wigfield_expectancy-value_2000}. In other words, students are more likely to complete a task that they expect to do well in and whose outcomes they think are valuable. 

There are four components of achievement values: attainment value (i.e. how important an individual thinks it is to do well on a task for their sense of self), intrinsic value (i.e. enjoyment of the task), utility value (i.e. how an activity is useful for an individual's future goals), and cost (i.e. effort, time, loss of valued alternatives and perceived cost of failure) (see Eccles et al. \cite{eccles_expectancies_1983}, and Wigfield and Eccles \cite{wigfield_expectancy-value_2000}  for a more detailed explanation).  

In terms of expecting to do well, the AI-generated hints on course assignments could increase student expectations of success on the assignments. In terms of value, scaling extra credit points for the supplemental math assignments could increase the utility value of those assignments, especially for students that did not do well on the midterm exams. And given that the AI-generated hints are integrated into the homework platform in Kudu and don't take much additional time/effort, the cost is minimal. 

AI-generated hints also reduce another aspect of the cost of getting help. Students often experience social barriers, including the fear of negative evaluation, that hinders them from asking instructors, teaching assistants, and/or peers for help \cite{jack_2015,Yee_2016}. However, such social barriers do not exist when using AI tools. AI-generated hints might be particularly helpful for students from marginalized backgrounds that are severely underrepresented in the classroom and/or prone to stereotype threat, which is the term for ``when members of a stigmatized group find themselves in a situation where negative stereotypes provide a possible framework for interpreting their behavior, the risk of being judged in light of those stereotypes can elicit a disruptive state that undermines performance and aspirations in that domain" \cite{spencer2016}.

\subsection{Research questions}

We explored students' use of optional supports, including supplemental math materials and AI-generated homework hints, and the effects on their exam performance. We were also interested in examining whether these optional supports are used equitably by students with different social identities and thus contribute towards more equitable student outcomes. Specifically, we sought to answer the following research questions (RQs):

\begin{itemize}
    \item RQ1: To what extent is students’ use of optional supports (specifically supplemental math assignments and AI-generated hints on homework problems) associated with gender, racial, and educational privilege?
    \item RQ2: Is using optional supports associated with an increase in student exam scores?
    \item RQ3: Is the use of optional supports associated with student grades similarly for students across a range of social identities? 
\end{itemize}

\subsection{Positionality of the authors}
As researchers and educators, our backgrounds and experiences shape the questions we ask and how we approach this work. 

Y.L. is a graduate student in the Department of Physics and Astronomy at UCLA. He is working towards his PhD degree in physics and has been Teaching Assistant for multiple introductory and upper-level undergraduate physics courses. Y.L. identifies as an Asian male.

K.S. is an Associate Director at UCLA’s STEM center for teaching and learning. She is a biologist and STEM education researcher with an emphasis on equity in STEM. KS identifies as a South Asian woman.  

A.K. is a Professor of Physics and Astronomy with twenty five years of experience in innovative teaching. He contributed to development of learning materials and digital tools incorporating AI and enabling \textit{active learning} which are used by tens of thousands of students at UCLA and other universities. A.K. led the development of supplemental mathematics materials and contributed to the development of AI tools that are used in this study.  He was the instructor for the courses included here. A.K. identifies as a White male. 

S.S. is a Senior Associate Director of UCLA’s STEM center for teaching and learning. She is a STEM education researcher who has taught introductory physics courses for life sciences majors at UCLA, but has not taught in the series for physical sciences majors described in this paper. In her teaching and education research, she has administered and interpreted the physics Force Concept Inventory. In terms of the identities described in this work, she identifies as a White female.

E.H.S. is a Professor of Physics at UC San Diego with thirty years of experience in interactive teaching. As a long-serving dean and chief academic officer, she has supported inclusive, experiential, and interdisciplinary pedagogy alongside education research across all academic fields. E.H.S. identifies as a White female.

\section{Methods}

\begin{figure*}[ht!]
    \centering
    \includegraphics[width=0.9\textwidth]{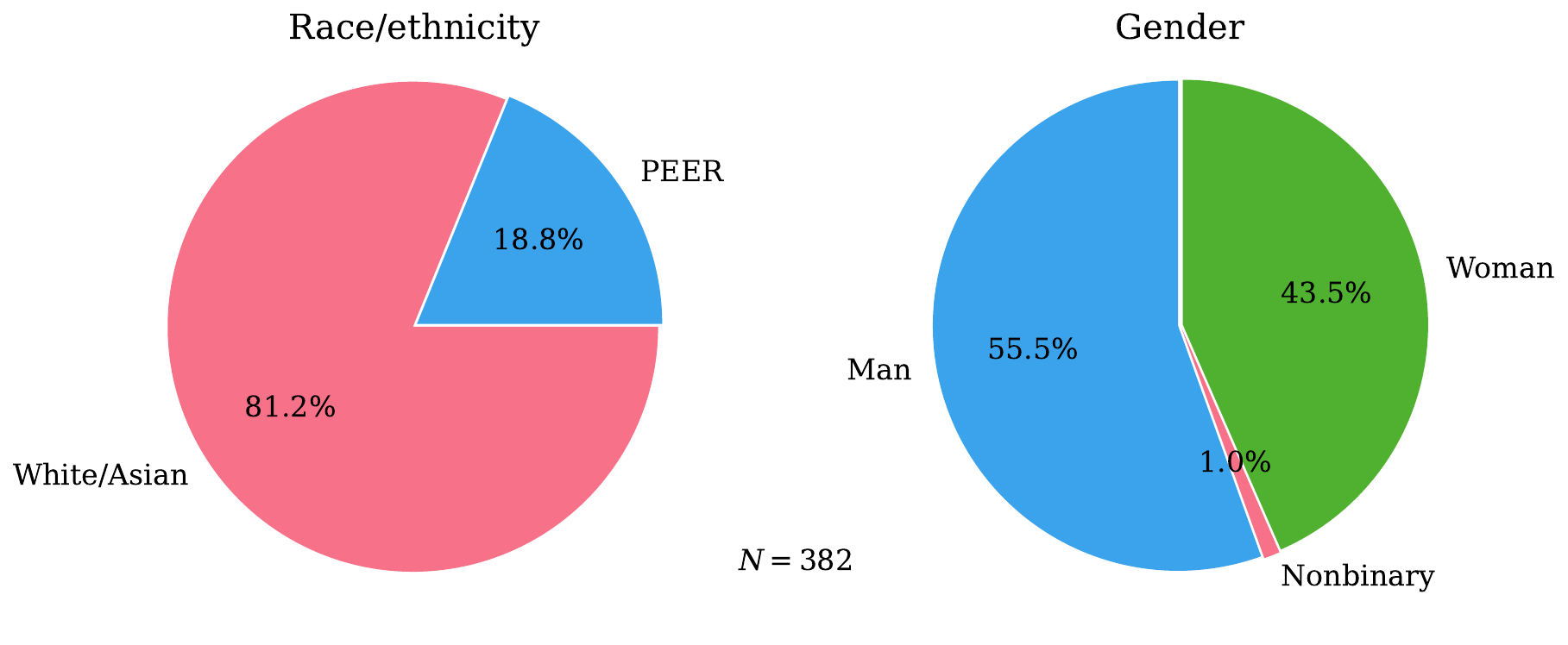}
    \captionsetup{width=\textwidth}
    \caption{Demographic composition of the participants in the study. Black/African American, Hispanic/Latine, and Native American students are grouped into the PEER (persons excluded from STEM because of their ethnicity or race) category.}
    \label{demographic}
\end{figure*}

The study was carried out at a large public university in the Southwest United States. The students participating in the study were enrolled in one of two sections (hereafter referred to as section A and B) of a calculus-based introductory physics course on Mechanics, typically the first in a series of General Physics courses. Both sections were taught by the same instructor. We obtained demographic data from institutional records. All study protocols were approved by the UCLA IRB. The basic demographic composition of the participants is summarized in \figref{demographic}. To protect student privacy and identifiable information, we grouped Black/African American, Hispanic/Latine, and Native American students together for analyses because of small cell sizes. Following the recommendation from Asai, we call this grouping "PEERs" i.e. persons excluded from STEM because of their ethnicity or race, because this term acknowledges that historical exclusion of people based on their race/ethnicity has been systematic and intentional \cite{asai2020excluded}.

\subsection{Study Design}

There were two forms of optional supports offered in the course: 1) supplemental math materials where students were encouraged to complete practice problems on a set of math skills that are related to mechanics, and 2) AI-generated hints for the physics homework questions. 

We adopted a quasi-experimental design to assess the association between use of optional supports and student exam performance. All supplemental materials and homework questions assigned to sections A and B were the same. 
The course was divided into three periods with one exam conducted at the end of each period. While math materials were available to students in both lectures, AI-generated hints and extra credit for completing the math materials were only available to one of the two sections during each period (except in Period 3). The detailed intervention schedule is listed in \figref{schedule}. 

To measure students' prior preparation, we administered two concept inventories at the beginning of the course to measure students' knowledge of Newtonian mechanics i.e. the Force Concept Inventory (FCI) \cite{hestenes1992force} and Calculus Concept Inventory (CCI) \cite{epstein2007development}. Throughout the three periods of this study, we collected data on student homework scores, use of AI-generated hints on the homework problems, and completion of supplemental math materials. We combined this data with demographics to address RQ1 (see \secref{finding1}). We then performed multiple linear regression with student exam performance as the outcome variable. This allowed us to quantitatively investigate RQ2 and RQ3 (see \secref{finding2a}, \secref{finding2a} and \secref{finding2b}).

\subsection{Optional supports}

The math skills relevant to students' success in introductory physics courses span a broad range. However, the key components generally involve mastery of vectors and familiarity with calculus-based tools such as derivatives and integration. This was the motivation behind the design of the supplemental math materials adopted in this course, which were divided into four chapters: 
\begin{itemize}
    \item Material 1: Vectors
    \item Material 2: Derivatives
    \item Material 3: Integrals
    \item Material 4: Multiple integrals
\end{itemize}
Example questions from each topic are shared in \appref{App:math}. At the beginning of the course, the instructor introduced the four supplemental materials together with the possible extra credit points associated with completing them. The materials were then assigned at different stages of the course. The release schedule was designed to match the ongoing course content while making sure students have abundant time to complete the materials prior to each exam.

\begin{figure*}[ht!]
    \centering
    \includegraphics[width=0.9\textwidth]{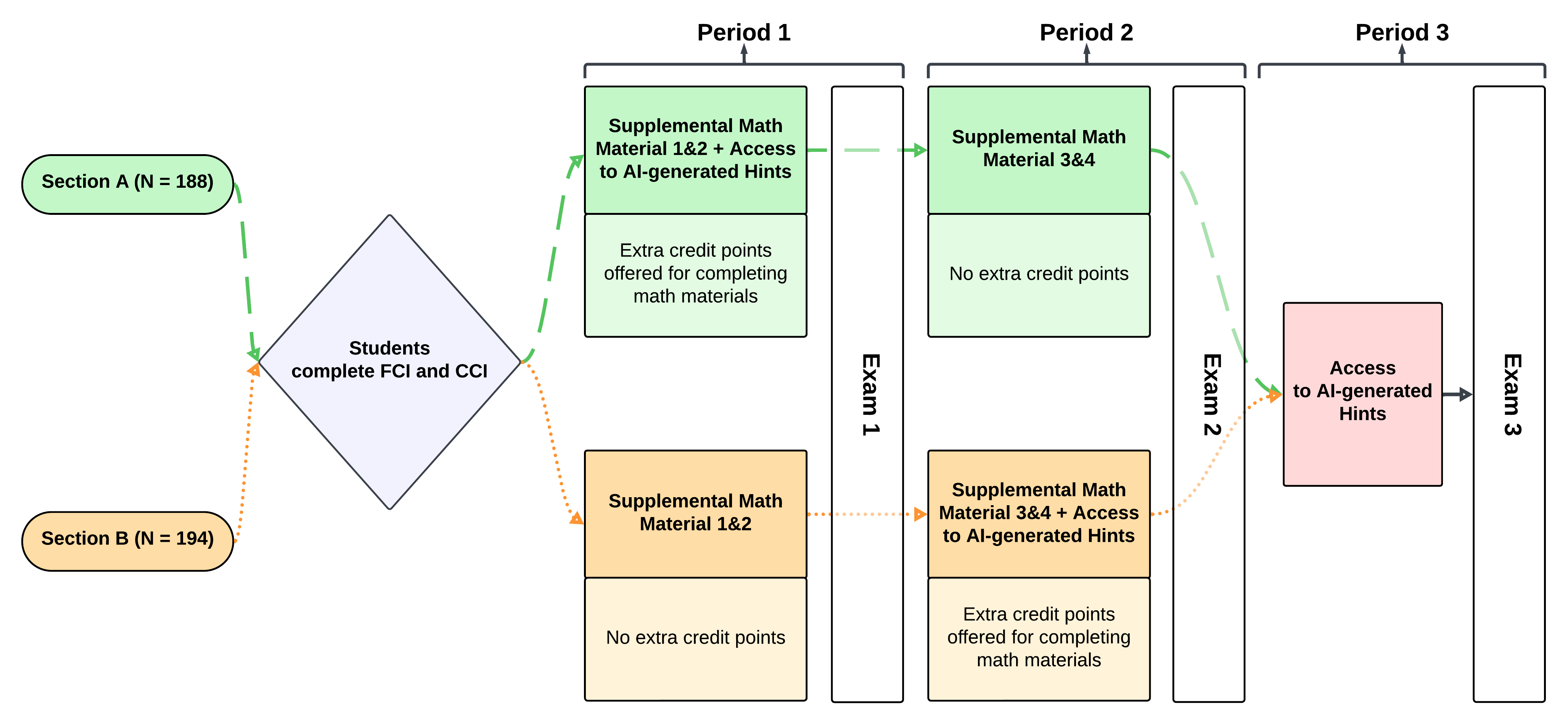}
    \captionsetup{width=\textwidth}
    \caption{Research design and the schedule of interventions.}
    \label{schedule}
\end{figure*}

Supplemental math materials were accessible to students in both lectures A and B, but incentives for completing them (in the form of extra credit points) were provided at different stages of the course in the two lectures (see \figref{schedule}). Furthermore, we adopted a formula for applying extra credit so that students who perform better in the exam would gain less by completing the materials. Specifically, the extra credit was determined by both the fractional score $X$ on the exam and the fractional score $Y$ on the supplemental material, according to the following formula:
\begin{equation}
    \mathrm{Extra\ credit} = 0.35 \, Y \cos \left (\frac{\pi}{2}X \right ).
    \label{extracredit}
\end{equation}
Note that the due date for the supplemental math materials was prior to the exam, so students did not know exactly how many points the supplemental materials would be worth to them. In other words, students might not have an accurate sense of the utility value of completing the math materials. 

The AI-generated hints are qualitative explanations related to physics homework questions and often include key ingredients to solve problems.   When a student requests a hint, the AI engine is provided with the text of the entire chapter and the text of the assignment, along with a carefully designed prompt instructing the AI to refrain from giving the full solution to the problem, while pointing the student in the right direction. An example of the AI-generated hints is demonstrated in \appref{App:AIhint}. We have carefully inspected the quality of such hints and made sure that they are both relevant and beneficial to build up students' problem solving skills in this course. When AI-generated hints are available, students have unlimited attempts to use them for each question so that they can explore the most useful version.

\subsection{Exam question evaluation}\label{sec:exameval}

\begin{table*}[ht!]
\centering
\begin{tabular}{P{1.5cm}|P{2cm}|P{2cm}|P{2cm}|P{2cm}|P{2cm}}
\hline \hline
& ISI & RI: material 1 & RI: material 2 & RI: material 3 & RI: material 4\\
\hline
Exam 1 & 0.0 & 0.6 & 0.2 & 0.2 & 0.0\\
Exam 2 & 0.4 & 0.2 & 0.0 & 0.2 & 0.0 \\
Exam 3 & 0.0 & 0.8 & 0.2 & 0.0 & 0.0 \\
\hline \hline
\end{tabular}
\caption{Exam evaluation summary according to the proportion of exam questions that are isomorphic to homework questions, the Isomorphism Index (ISI), and the proportion of questions that require the same math skills as a given math material, the Relevance Index (RI).}
\label{index}
\end{table*}

There were three exams used for summative assessment in the course, two midterms and a cumulative final. To disentangle the effect of AI-generated hints versus supplemental math material, we evaluated all exam questions to assess their alignment with the math included in the supplemental math assignments and Physics homework problems. The evaluation process was carried out as follows: 1) we selected a subset of homework questions that were relevant to the exam questions and created a question pool for each exam. 2) For each exam question, we determined if there was an isomorphic question (i.e. a pair of questions that require the same set of concepts and skills to solve, with differences only at the surface level, such as the provided numerical values or the problem context) in the question pool. 3) For each exam question, we also matched it to one or more supplemental math materials according to the math skills required to solve it. An example of a pair of isomorphic questions is given in \appref{App:isomorphic}. This evaluation was performed by Y.L. who has a bachelor's degree in physics and has taught undergraduate physics courses. 

To quantify our evaluation results, we assigned an Isomorphism Index (ISI) to each exam, defined as the proportion of exam questions that are isomorphic to homework questions. In addition, for each supplemental math material, we calculated the Relevance Index (RI) of the exam by the proportion of questions that require the same math skills as a given math material. The evaluation result is summarized in \tabref{index}. 

\subsection{Regression models}

\begin{table*}[ht!]
\centering
\begin{tabular}{lc}
\hline\hline
Variable & Description\\
\hline
Homework score    & Fractional homework score between 0 and 1 \\
AI hints usage    & Fractional problems that students used AI hints on\\
Math supplement material& Completion (binary) of material 1-4 \\
Force concept inventory  & Force inventory test score (z-score) \\
Calculus concept inventory    &  Calculus inventory test score (z-score)  \\
Gender & Man (reference) and URG \\
Race/ethnicity & WA (reference) and PEER  \\
\hline\hline
\end{tabular}
\caption{List of variables used in the regression models.}
\label{regression variables}
\end{table*}

We used multiple linear regression to assess the association between optional supports and student exam performance. Student exam scores were used as the outcome variable. To standardize the data, we converted the exam scores into z-scores with zero mean and standard deviation 1, according to the following transformation
\begin{equation}
    \text{z-score} = \frac{\text{raw exam score} - \mu}{\sigma},
\end{equation}
where $\mu$ and $\sigma$ are the mean and standard deviation of each exam. The following variables were used as predictors: 1) prior preparation measured by the concept inventories and converted to z-scores, 2) homework scores, 3) optional support usage, including use of AI-generated hints and completion of supplemental math material, and 4) demographics, specifically gender and race/ethnicity. For each exam, we used the question pool from the exam evaluation stage (\secref{sec:exameval}) and quantified AI-generated hints use as the proportion of problems that a student used AI-generated hints on among all the problems in the question pool. We note that each exam has a separate question pool, so that AI-generated hints use is only measured on the relevant (but not necessarily isomorphic) homework questions. The race/ethnicity and gender variables were converted to binary variables to protect student privacy given some small cell sizes in the disaggregated data. White/Asian (WA) students were used as reference to PEER (persons excluded due to race/ethnicity including Black/African American, Hispanic/Latine, and Native American students, and men were used as reference to students with underrepresented gender identities (URG: women and non-binary). The details of all variables used in the model are listed in \tabref{regression variables}.

The effect of a variable in the regression model is reflected through its coefficient and its $p$ value. For RQ3, it is crucial to take into account the effect of using optional support on different student groups. Therefore, we added an additional interaction term in the model, i.e. the product of two variables of interest. For example, we were interested in the interaction between completing math supplemental materials ($x_{\rm supp}$) and student's race/ethnicity group ($x_{\rm re}$). So the regression model took the form
\begin{equation}
    y = \cdots + \beta_{\rm re} x_{\mathrm{re}} + \beta_{\rm supp} x_{\mathrm{\rm supp}} + \beta_{int} x_{\mathrm{re}} x_{\mathrm{\rm supp}},
    \label{int_regression}
\end{equation}
where the dots indicate regression terms from other variables. When one of the variables in the interaction term ($x_{\mathrm{re}}$ here) is binary, the interaction coefficient has a simple interpretation: if the binary variable indicates the reference group (taking value 0), then the effect of the second variable is solely expressed in terms of its coefficient. However, when the binary variable takes the value 1, the effect of the second variable comes from the combination of its coefficient and the interaction coefficient. Therefore, in our example, $\beta_{int}$ indicates the difference between the effect of completing math supplemental materials between the race/ethnicity groups.

When we tested the assumptions for multiple linear regressions, we found that there is heteroscedasticity in our data. Therefore, to reliably estimate the $p$ values, we adopted the heteroscedasticity-consistent covariance 'HC3' \cite{mackinnon1985some} when performing the regression. In addition, some variables in \tabref{regression variables} are strongly correlated. As an example, Math supplement material 1 and 2 are positively correlated as they are incentivized together. We remedied the multicollinearity issue by dropping some of the variables when necessary (see \secref{finding2a} for more details).

To quantify factors that are associated with students' completion of math materials, we performed a logistic regression. Here, the outcome variable was whether or not students completed a given supplemental math assignment, and predictor variables included prior preparation, demographics, and which section students were enrolled in. 

All data analyses were completed using \lstinline|Pandas| \cite{reback2020pandas, mckinney-proc-scipy-2010}, \lstinline|Matplotlib| \cite{Hunter:2007} and \lstinline|Seaborn| \cite{Waskom2021}.

\section{Results}

188 students in Section A and 194 students in Section B participated in this study. After converting the FCI and CCI results into z-scores, we compared the distribution from Section A (FCI: mean = -0.04, standard deviation (SD) = 1.02, CCI: mean = -0.08, SD = 1.04) and Section B (FCI: mean = 0.04, SD = 0.99, CCI: mean = 0.07, SD = 0.96). Furthermore, Welch’s t-test on the FCI and CCI scores between the two sections showed no significant difference in prior preparation ($p = 0.41\ \mathrm{and}\ 0.15$, respectively). However, we found significant differences in prior physics preparation across gender ($t= 5.6$, $p<0.001$) and race/ethnicity ($t = 4.5$, $p<0.001$), and similarly in math background across gender ($t= 4.1$, $p<0.001$) and race/ethnicity ($t = 3.4$, $p<0.001$).

 Male students scored an average of 0.24 (SD = 1.02) on the FCI compared to URG students who had an average of -0.30  (SD = 0.90). WA students scored an average of 0.10 (SD = 0.99), while PEER students had mean of -0.44 (SD = 0.91). 

On the CCI, male students scored an average of 0.19  (SD = 0.93) while URG students had mean of -0.23 (SD = 1.04). WA students scored an average of 0.09 (SD = 0.98) compared to PEER students who had mean of -0.37 (SD = 1.03).

\subsection{Finding 1: Incentives are associated with higher completion rate and reduced disparities in completion of math supplemental material for PEERs}
\label{finding1}

\begin{figure}[ht!]
    \centering
    \includegraphics[width=0.9\columnwidth]{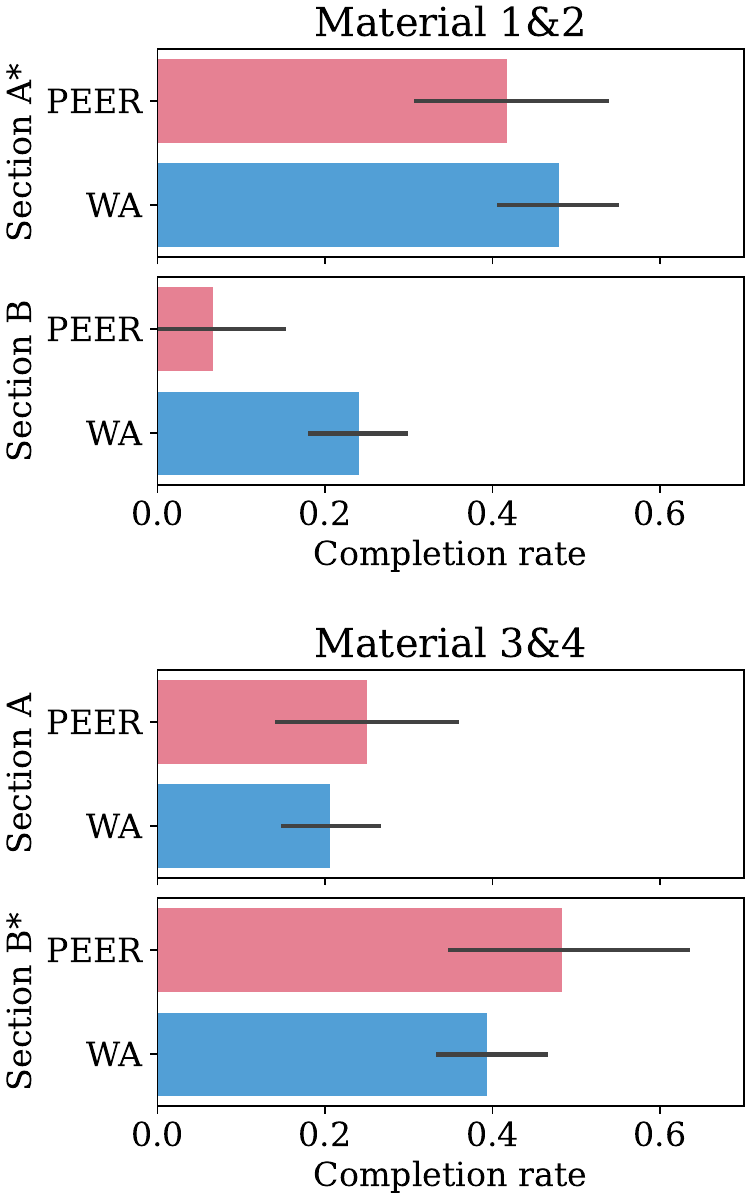}
    \caption{Comparison of supplemental math material completion rate with different incentive availability across race/ethnicity groups. Stars (*) indicate incentives are applied to the corresponding section and the black lines mark $95\%$ confidence intervals. PEER stands for persons excluded due to race/ethnicity and includes Black/African American, Hispanic/Latine, and Native American students, WA stands for White/Asian American.  }
    \label{completion_peer}
\end{figure}

On average, the completion rates for supplemental math assignments when unincentivized were around $20\%$ and offering extra credit increased the completion rate by about $20\%$. Furthermore, when incentives were not offered, we observed a significantly low completion rate for PEERs in Section B, compared to WA students (upper panel in \figref{completion_peer}). But this disparity was alleviated when extra credit was offered. Our full logistic regression results are summarized in \appref{otherresult}. 

\begin{figure}[ht!]
    \centering
    \includegraphics[width=0.9\columnwidth]{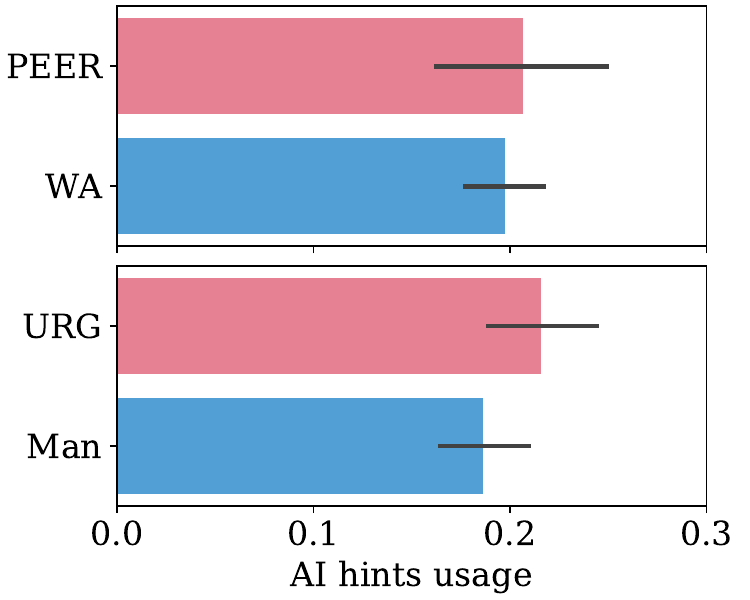}
    \caption{AI hints usage rate across demographic groups. The black lines indicate $95\%$ confidence intervals. PEER stands for persons excluded due to race/ethnicity and includes Black/African American, Hispanic/Latine, and Native American students, WA stands for White/Asian American. URG stands for under-represented gender and includes women, non-binary students, and other gender minorities}
    \label{aihints}
\end{figure}

We observed no difference in the use of AI-generated hints across gender and race/ethnicity groups \figref{aihints}. Most of the students explored the AI-generated hints feature in Kudu: when available, $88\%$ of the students used AI-generated hints on at least 1 homework question (among the problems that we found to match the exam questions), and $57\%$ students used AI-generated hints on at least 5 questions. Depending on the section, the question pool size is around 40. On average, students used AI-generated hints on at least $20\%$ of the questions.

\subsection{Finding 2a: Completing supplemental math assignments was associated with improved student exam performance, especially when math material was most relevant to exam content}
\label{finding2a}

\begin{table*}[ht!]
\centering
\begin{tabular}{lccc}
\hline\hline
& Coefficient & Standard error  & $p$ value\\
\hline
\textbf{Intercept}  &      -2.97  &        0.34     &         $<0.001$       \\
Gender: URG&      -0.04  &        0.07     &         0.611       \\
(Reference = Man)&&&\\
\textbf{Race/ethnicity}: PEER&      -0.22  &        0.10    &         0.030        \\
\textbf{(Reference  = WA)} &&&\\
\textbf{Homework score}    &       3.16  &        0.35     &         $< 0.001$       \\
AI hints usage         &       0.24  &        0.21     &         0.240       \\
\textbf{Math supplement material 1} &       0.24  &        0.07      &         0.001      \\
Math supplement material 3 &       0.05  &        0.07     &         0.444      \\
\textbf{Force concept inventory}&       0.31  &        0.06    &         $<0.001$        \\
\textbf{Calculus concept inventory}&       0.12  &        0.06      &         0.029       \\
\hline\hline
\end{tabular}
\caption{OLS regression model for Exam 3 score. Variables with statistical significance are marked in bold.}
\label{exam3ols}
\end{table*}

The regression model for Exam 3 is shown in \tabref{exam3ols} and we list the results for Exam 1 and 2 in \appref{otherresult}. None of the exam problems used the math skills covered in supplemental math material 4, thus we excluded this variable from the model. Additionally, there was a strong correlation between math material 1 and 2 since they were incentivized together, therefore we dropped the math material 2 variable from our models. We did not find a significant correlation between gender and exam performance. However, PEER students had significantly lower exam scores compared to WA students. The correlation between demographics and exam performance was consistent across all exams. In addition, prior preparation in physics/math had a significant association with student exam scores, which is consistent with the findings in \refref{PhysRevPhysEducRes.13.020137}.

\begin{figure*}[ht!]
    \begin{subfigure}{0.3\textwidth}
        \includegraphics[width=\textwidth]{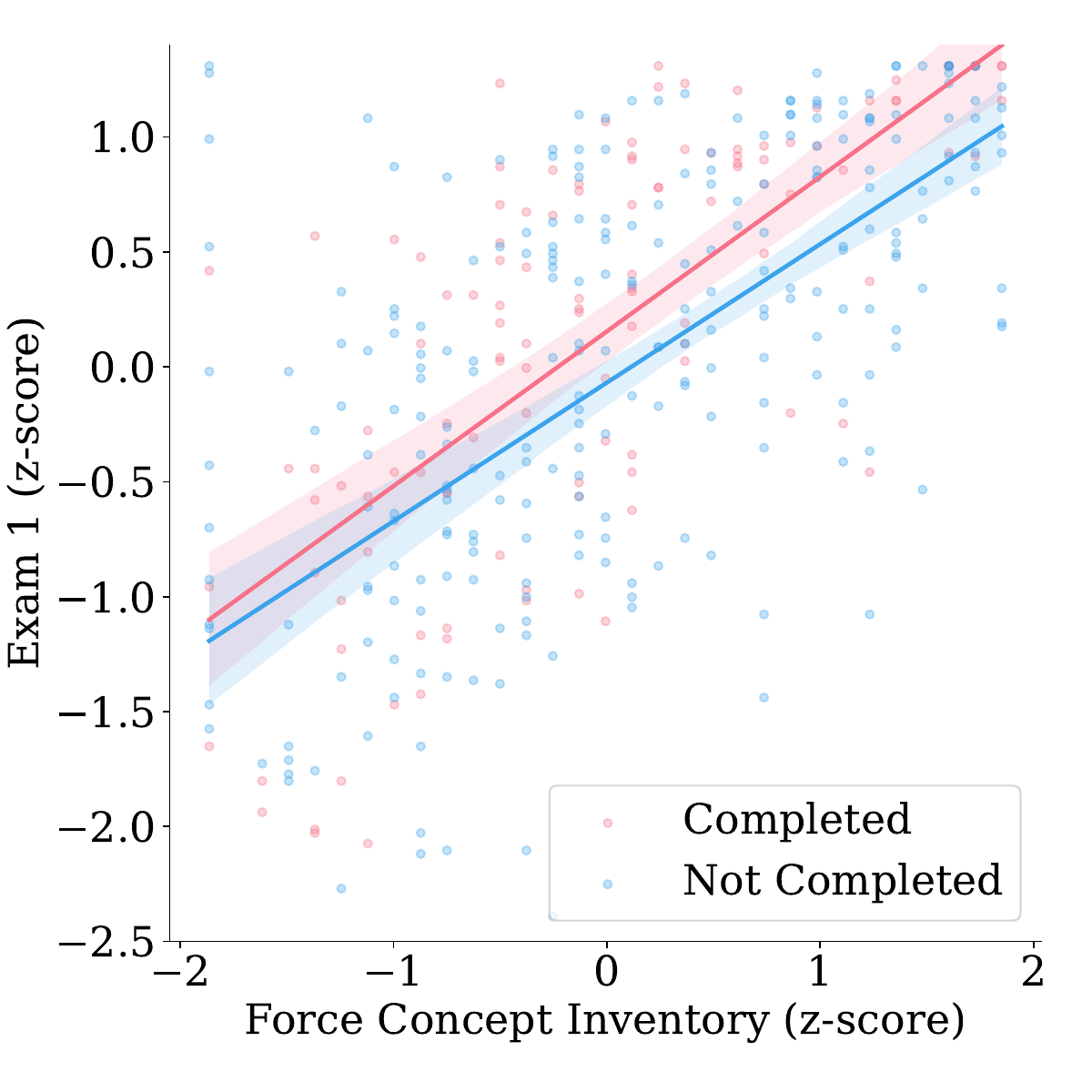}
        \caption{$\rm RI = 0.6$}
    \end{subfigure}
    \begin{subfigure}{0.3\textwidth}
        \includegraphics[width=\textwidth]{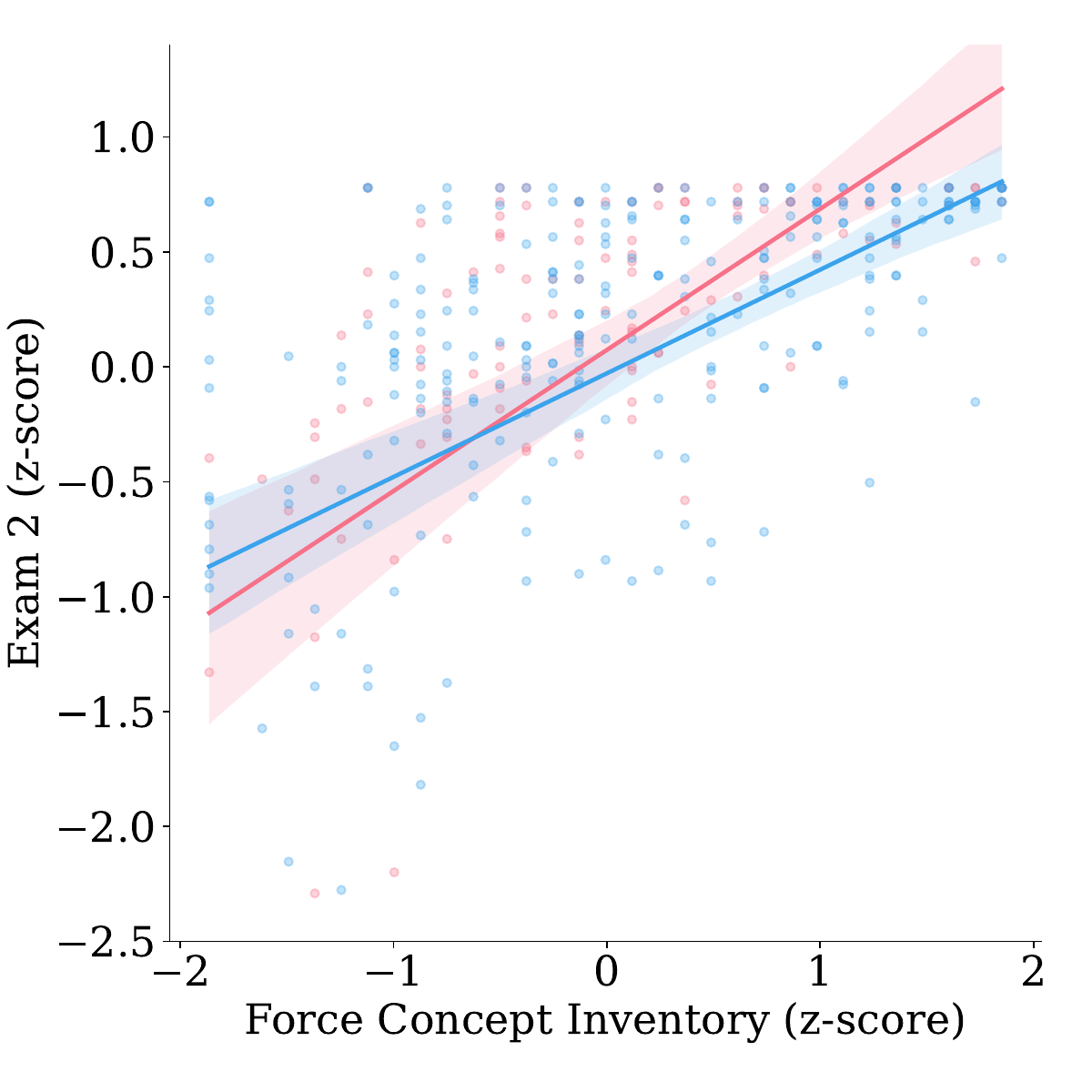}
        \caption{$\rm RI = 0.2$}
    \end{subfigure}
    \begin{subfigure}{0.3\textwidth}
        \includegraphics[width=\textwidth]{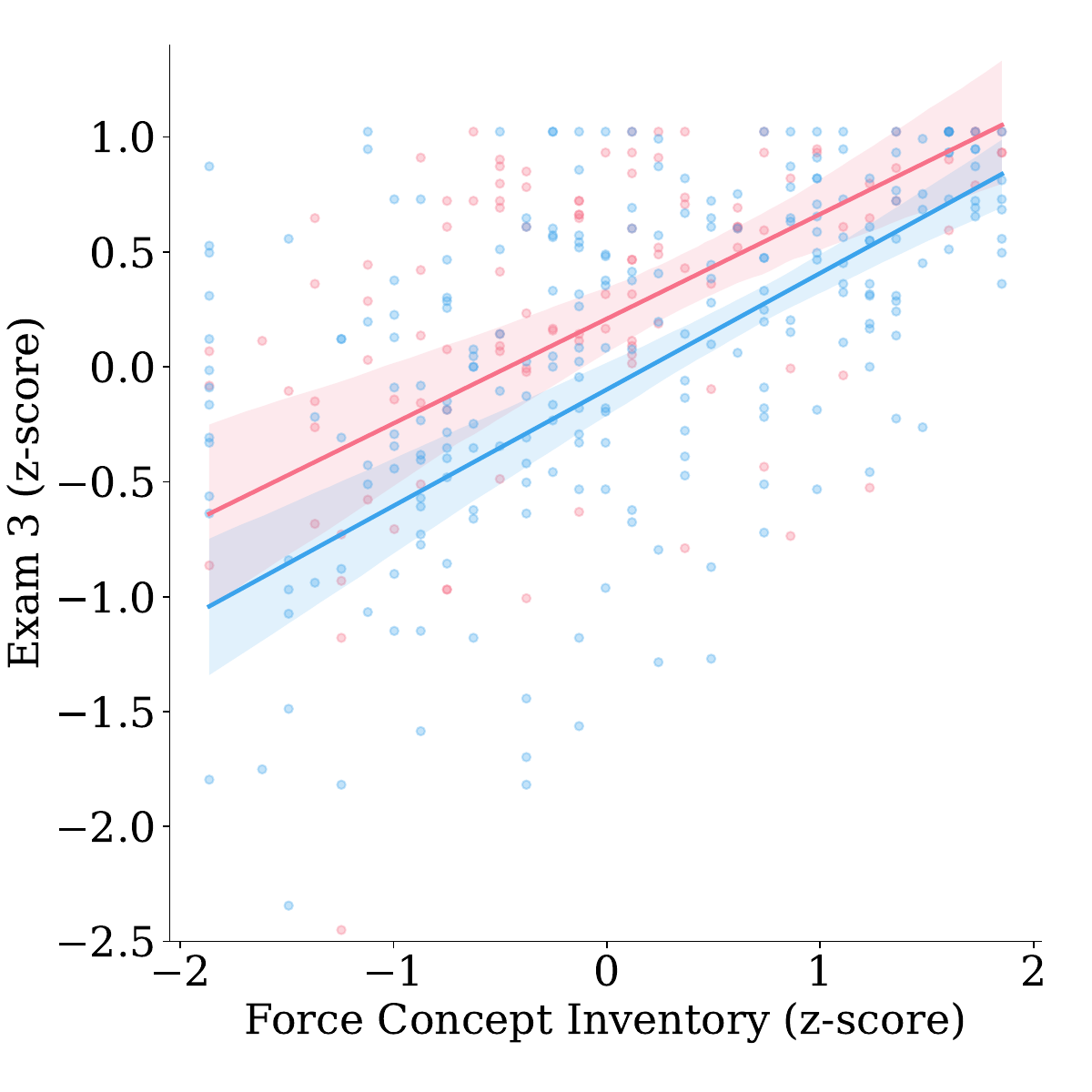}
        \caption{$\rm RI = 0.8$}
    \end{subfigure}
    \captionsetup{width=\textwidth}
    \caption{Correlation of student exam performance with Force Concept Inventory score, separated by supplemental math material 1 completion. The colored bands mark regions with $95\%$ confidence interval (after taking into account other variables in the model).}
    \label{regression_fit}
\end{figure*}

Notably, the coefficient of supplemental math material 1 completion is positive and statistically significant, showing that additional practice on vectors is indeed positively associated with better exam performance. We did not find a significant association between exam performance and other supplemental math assignments. This is because the topics of other supplemental math assignments (derivatives and integrals) are not relevant to the Introductory Mechanics course we studied. However, those topics are relevant for the subsequent courses in the Introductory Physics series. 

When we compared the relevance of material 1 indicated by our Relevance Index (RI, i.e. proportion of physics exam questions that used the math skills on a given math assignment) and its regression coefficient across all exams, we found that as RI increases, the regression coefficient becomes not only significant but also larger (see \tabref{exam3ols}, \tabref{exam1ols}, and \tabref{exam2ols}). This effect is visually demonstrated in \figref{regression_fit}, where we plotted the best fit lines of exam score versus FCI score for students that completed supplemental math material 1 and those that did not. We observed a clear separation of the fitted lines for Exam 1 ($\rm RI = 0.6$) and Exam 3 ($\rm RI = 0.8$) with larger separation for Exam 3, but this is absent for Exam 2, which only has a RI of 0.2.

\subsection{Finding 2b: Using AI-generated hints is associated with improved student exam performance when exam content is aligned with the homework questions}
\label{finding2b}

\begin{figure}[ht!]
    \centering
    \includegraphics[width=0.9\columnwidth]{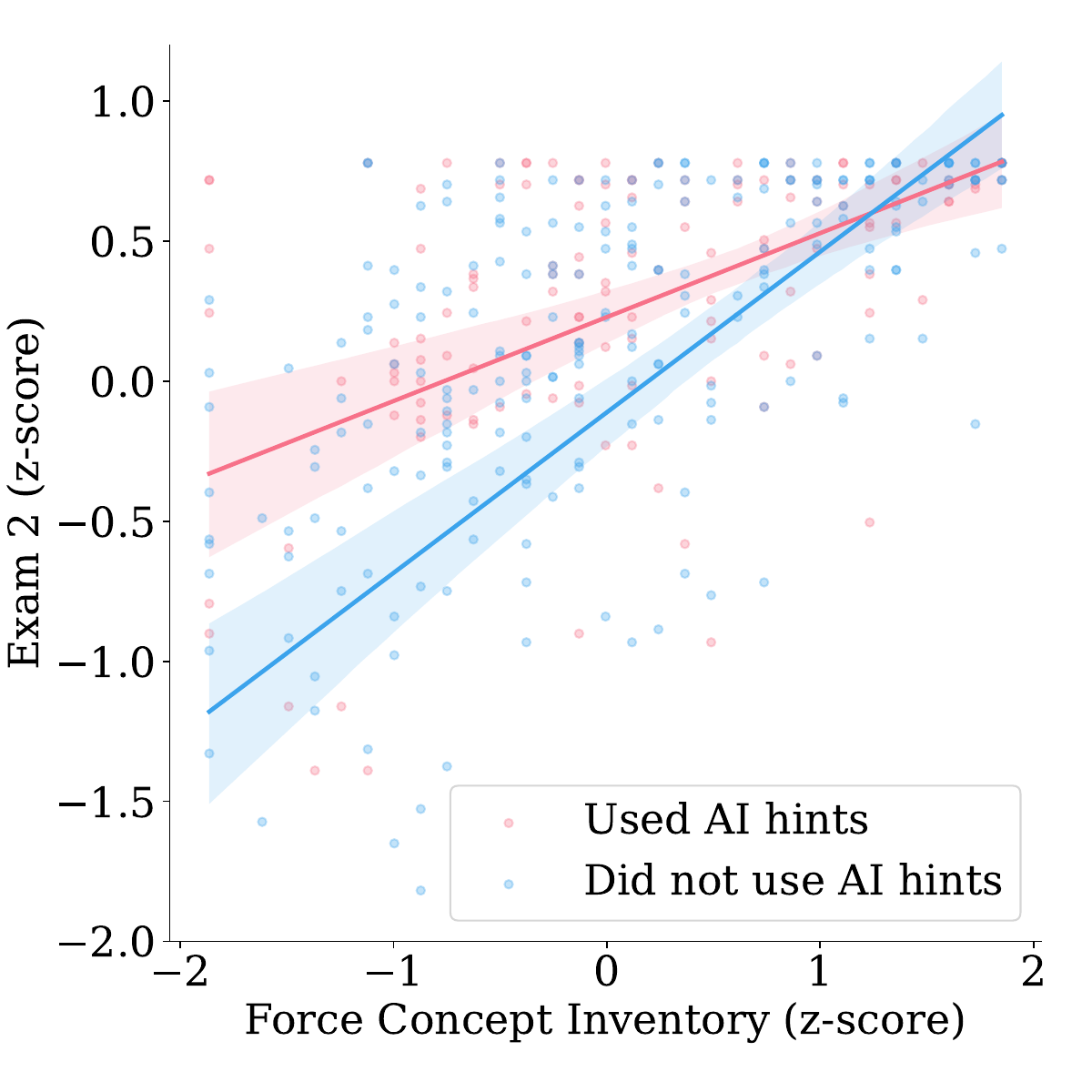}
    \caption{Exam 2 score versus Force Concept Inventory score, separated by AI hints usage. The colored bands indicate $95\%$ confidence region.}
    \label{exam2ai}
\end{figure}

We found a statistically significant association between Exam 2 score and AI hints usage, as demonstrated in \figref{exam2ai} and \tabref{exam2ols}. Interestingly, the linear fit result in \figref{exam2ai} suggests that AI-generated hints might benefit students who are less prepared to a larger extent. To examine this effect, we added an interaction term between the AI hints usages and FCI variables, with the updated regression result shown in \tabref{exam2int}. Indeed, our observation is reflected in the negative coefficient of the interaction term, although not statistically significant. However, we did not find a significant association between student exam performance and AI hints usage for Exams 1 and 3, possibly due to the low relevance between these exams and the homework questions, as indicated from the ISI in \tabref{index}.

\subsection{Finding 3: Completing supplemental math assignments might be associated with lower disparities in exam scores between WA students and PEERs}
\label{finding3}

When we added an interaction between competing supplemental math material 1 and race/ethnicity in our regression model, we found that the interaction coefficient was positive with a large effect size (coefficient of 0.38, compared to a coefficient of 0.32 for the Force Concept Inventory) and a $p$ value of 0.081 (see \tabref{exam3int}). Although the  $p$ value does not reach the conventional threshold at 0.05, the interaction term still suggests that completing supplemental math materials might benefit PEER students more. 

The regression coefficients suggest that for two WA students with an average homework score and concept inventory score, the one who completed supplemental math assignment 1 would increase their exam score by about a fifth of a standard deviation. However, for two PEER students who had an average homework and concept inventory score, the one who completed the supplemental math assignment 1 would increase their exam score by more than half of a standard deviation, with contributions from both the math supplemental material 1 coefficient and the interaction term coefficient. Therefore, our result provides some tentative evidence that the additional math practice is not only beneficial to the entire class, but it also improves the exam performance of PEERs' students even more than others. 

\begin{table*}[ht!]
\centering
\begin{tabular}{lccc}
\hline\hline
  & Coefficient & Standard error  & $p$ value\\
\hline
\textbf{Intercept}   &      -2.94  &        0.34     &       $<0.001$\\
Gender                     &      -0.03  &        0.07    &        0.679 \\
(Reference = Man) &&&\\
\textbf{Homework score}    &       3.14  &        0.35     &      $ <0.001$ \\
AI hints usage          &       0.24  &        0.21     &       0.245 \\
Math supplement material 3 &       0.06  &        0.07     &       0.414 \\
\textbf{Force concept inventory}  &       0.32  &        0.06     &      $ <0.001$ \\
\textbf{Calculus concept inventory}     &       0.12  &        0.06     &       0.023 \\
\textit{\textbf{Race/ethnicity}}             &      -0.32  &        0.12     &       0.006 \\
\textit{\textbf{(Reference = WA)}} &&&\\
\textit{\textbf{Math supplement material 1}} &       0.18  &        0.08     &       0.020 \\
\textit{Race/ethnicity : Math material 1 }     &       0.38  &        0.22     &       0.081 \\
\hline\hline
\end{tabular}
\caption{Regression model on Exam 3 with interaction term. The last row corresponds to the interaction term. Statistically significant variables are marked in bold and the relevant interaction variables are marked in italics. The positive coefficient for the interaction term indicates that the positive effect of math material 1 is stronger for PEERs.}
\label{exam3int}
\end{table*}

\section{Discussion}
We offered two kinds of optional supports to students in an Introductory Physics course: supplemental math assignments and AI-generated hints. We found that completing the math assignments and/or using AI-generated hints was associated with better exam performance on the exams that were most related to these supports (\figsref{regression_fit} {exam2ai}). 

Previous studies have shown that online supplemental math assignments can be effective in improving student performance in Physics courses \cite{PhysRevPhysEducRes.13.010122, PhysRevPhysEducRes.13.020137}. In this study, we tried a unique scaled incentive structure for free online supplemental math assignments to increase the completion rate, especially among students that need more support with mathematical skills. With this incentive structure, we were able to achieve higher completion rates among students who might need more support in the course compared to students who are well equipped to perform well in the course, as indicated by students' incoming FCI scores. In other words, when incentivitized with the scaled extra credit, about 54 percent of students that scored below the class mean on the FCI completed the first supplemental math assignment compared to only 37 percent of students that scored at or above the class mean. By contrast, without the incentive, only 19 percent of students that scored below the class mean on FCI completed the first supplemental math assignment, compared to 24 percent of students that scored at or above the mean. When we broke this down by race/ethnicity, we found that our scaled incentive structure was associated with higher completion rate among students from historically under-represented racial/ethnic backgrounds. The higher supplemental math completion rate among PEERs, combined with the improved performance among students who completed supplemental math assignment 1, suggests that offering these assignments along with a scaled incentive structure can decrease racial inequities in Physics courses.  

Students who complete math assignments might do better on the exams simply because they are more motivated to do well in the course, and/or because they tend to spend more time working on the course. If motivation and/or time on task were the main reasons for improved exam performance, then we would have seen improved exam performance regardless of the alignment between the exams and math materials. However, our results only show positive association between completing supplemental math assignment 1 and performance on exams 1 and 3 which have the highest proportion of aligned questions; we do not find such an association for exam 2 which had a low proportion of aligned questions. Similarly, we did not find a positive association between exam performance and completing math materials 3 and 4, which were not aligned well with any of the exams in this course. Note that materials 2, 3, and 4, are useful for downstream courses in the introductory Physics series, so students that completed these might perform better in those subsequent courses. We plan to explore this idea in future studies. 

Our study is one of the first to explore the impact of offering LLM (Large Language Model)-based AI-generated hints during homework assignments on student performance in exams in an Undergraduate Physics course. Given the differences in the quality of AI available in open-source compared to paid formats (GPT 3 vs 4 at the time of this study) \cite{yeadon_impact_2024, ding_students_2023}, differential access to AI could increase inequities in student performance on homework assignments and exams. In this course, high-quality AI-generated hints for homework problems were made available to all students through the course platform Kudu. We found that a large majority of students used at least one AI-generated hint and there were not significant differences in the use of AI-generated hints by gender and race/ethnicity. Moreover, there was a positive association between use of AI-generated hints and exam 2 performance, which was the exam most aligned with homework assignments. This suggests that high-quality generative AI tools, if made available to everyone, have the potential to support student learning equitably. In future studies, we plan to explore further students' reasons for using or not using AI hints and their experience with using these hints. 

\subsection{Limitations and Future Directions}
Although our interventions are based on the Situated  Expectancy-Value Theory of Achievement Motivation, we did not collect survey data on student's expectancy beliefs and values regarding the supplemental math assignments and AI-generated hints. We also did not collect data on student use of any other AI-powered tools or use of any other optional supports such as tutoring services. We plan to address these shortcomings in future studies. Another limitation of our study is that we were not able to disaggregate the race/ethnicity and gender data because of smaller sample sizes in some of the groups we have included in those categories. We recognize that combining these groups with different historical, social, and cultural context obscures patterns of racial and gender inequality \cite{bensimon2016misbegotten}. In future work, we plan to collect larger datasets that could allow us to diasggregate these data for analyses. 

A limitation of our intervention is that despite the scaled extra credit incentive structure, the completion rate for supplemental math assignments is still less than 60 percent. Integrating the supplemental math assignments into formal class time as a component of discussion section might further increase student participation.  Sharing the results of this study with future students may also be helpful in increasing the perceived utility value of the supplemental math assignments. 

In future studies, we also plan to collect some qualitative data using surveys and interviews to give us insights into the factors that shape students' decisions to use the optional supports offered in the course. This qualitative data will help us understand the perceived value of these optional supports and inform structural modifications to increase student uptake of these supports.

\section*{Acknowledgments}
We thank Warren Essey, Stuart Brown, and Joshua Samani for valuable discussions and input. We also thank Harrison Parker for designing the teaching intervention flow chart. This work was supported by UCLA's Teaching and Learning Center and by the UCLA Department of Physics and Astronomy.
\bibliographystyle{apsrev4-1}

\bibliography{main.bib}

%%%%%%%%%%%%%%%%%%%%%%%%%%%%%%%%%%%%%%%%%%%%%%%%%%%%%%%%%%%%%%%%%%%%%%%%%%%
\clearpage

\appendix

\appendixpage

\section{Sample supplemental math questions}\label{App:math}

The supplemental math materials contain both qualitative and quantitative questions covering high school to beginning college level math on vectors, derivatives and integrals. An example from each material is given below:

\medskip

\textbf{Material 1 (Vectors)}:

A two-dimensional vector has an $x$-component 
of 8.71 meters and a $y$-component of 5.43 meters. 
Calculate the angle (in degrees) that this two-dimensional vector makes with the positive $x$-axis. 

\medskip

\textbf{Material 2 (Derivatives)}:

Calculate the derivative of $3 \sqrt{x^5 + 1} + 5$ with respect to $x$.

\medskip

\textbf{Material 3 (Integrals)}:

A car antenna that is 0.726 m long (starting at $x=0$) has a density function $\rho(x) = e^{x/2}$ kg/m. Find the mass of the antenna. Hint: the mass of an antenna between 0 and $L$ is $m=\int_0^{L}\rho(x) dx$

\medskip

\textbf{Material 4 (Multiple integrals)}:

A thin disk of radius $R$ has a uniform mass density $\rho$. You are asked to find the gravitational force experienced by a point mass $m$ located at a distance $z$ above the center of the disk. Use the formula for the gravitational force due to a mass element $dM$ at distance $r$: $dF = G m/r^2 dM$. Which of the integrals represents the correct answer? 

\begin{enumerate}[label=\alph*)]
    \item $F=\int_0^R \frac{2 \pi G m \rho}{r^2+z^2} r d r$

    \item $F=\int_0^R \frac{4 \pi G m \rho}{r^2+z^2} r^2 d r$

    \item $F=\int_0^R \frac{4 \pi}{3} \frac{G m \rho}{r^2+z^2} d r$

    \item $F=\int_0^R \frac{4 \pi G m \rho}{\sqrt{r^2+z^2}} d r$

    \item $F=\int_0^R \frac{\pi G m \rho}{\sqrt{r^2+z^2}} r^3 d r$
\end{enumerate}

\section{Sample AI-generated hints}\label{App:AIhint}

The AI-generated hints are designed to give qualitative descriptions on the essential steps towards solving a problem. They are meant to \textit{guide} students in the problem solving process instead of completely giving away the answer. A typical example from Introductory Mechanics is demonstrated below:

\textbf{Problem}: ``A bird has a mass of 26 g and perches in the middle of a stretched telephone line. Determine the tension when both halves of the line have an angle of $5^{\circ}$ from the horizontal direction. Assume that each half of the line is straight.''

\textbf{Hint}: ``Start by separating the forces acting on the bird into its vertical and horizontal components. Keep in mind that the force due to gravity acts downwards and the tension in the wire acts along the wire at angle $\theta$ to the horizontal. Implement Newton's second law, which states that the sum of the forces must equal to zero as bird is in equilibrium. Solve the equations to find the tension.''

\section{Sample isomorphic questions}
\label{App:isomorphic}

In this section, we demonstrate a pair of exam and homework questions that are determined to be isomorphic in our exam question evaluation phase (\secref{sec:exameval}).

\textbf{Homework problem}: "Block 2 slides along a horizontal frictionless table as block 1 falls. The blocks are connected via a frictionless pulley. Find the speed of the blocks after they have each moved 2.0 m. Assume that they start at rest and that the pulley has negligible mass. Use $m_1 = 2$~kg and $m_2 = 4$~kg."

\textbf{Exam problem}: "Block 2 with mass $m_2 = 3$~kg slides along a horizontal table as block 1 with mass $m_1 = 2$ kg moves downward. The blocks are connected by a rope via a frictionless pulley; the rope and the pulley have negligible mass. Find the speed of the blocks after they have each moved 2.0 m starting from rest."

\section{Other results in the study}
\label{otherresult}

\begin{figure*}[h!]
    \centering
    \includegraphics[width=0.9\textwidth]{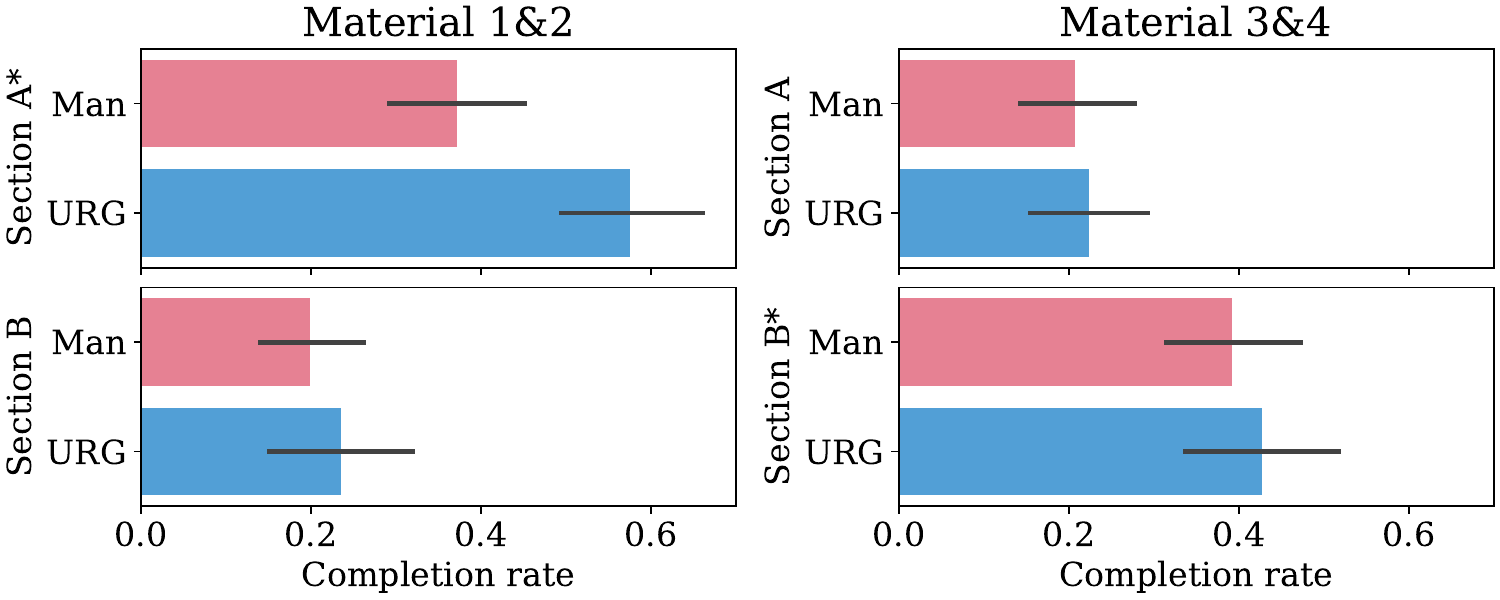}
    \caption{Comparison of supplemental math material completion rate across gender groups. Stars (*) indicate incentives are applied to the corresponding section and the black lines mark $95\%$ confidence intervals.}
    \label{completion_gender}
\end{figure*}

Here, we compile the rest of the results in our study not shown in the main text, including: 
\begin{enumerate}
    \item Comparison of math material completion rates across gender groups (\figref{completion_gender}).

    \item Logistic regression model predicting students completion of the math supplemental materials (Tab.~\ref{logistic12} and \ref{logistic34}).
    
    \item Ordinary linear regression model for Exam 1 and 2 performance (\tabref{exam1ols} and \ref{exam2ols}).

    \item Regression model for Exam 2 with additional interaction term (\tabref{exam2int}).
    
\end{enumerate}

Our logistic regression model predicts the binary variable that corresponds to the completion of supplemental math materials, where the prediction outcome can also be interpreted as the probability that a student will complete the material, given the explanatory variables of the model. In our model, we combined material 1 and 2 (similarly, for material 3 and 4) and define `completion' as completing both materials. From Tab.~\ref{logistic12} and \ref{logistic34}, we can clearly see the effect of offering extra credits on the completion status: the positive (negative) coefficient of the `Section' variable in material 1 and 2 (material 3 and 4) demonstrates that students are more likely to complete the material when extra credits are offered. Interestingly, student's gender and race/ethnicity play a role in the likelihood of completing math materials 1 and 2; however, they are not significant any more for math materials 3 and 4. This is also reflected in \figref{completion_peer}, and such observation could possibly be explained by a temporal effect: the initial incentives help students put a higher utility value on math materials. If students themselves are aware of the benefits of completing the math materials, then they are more likely to complete the math materials even if there is no extra credit associated with it, leading to the relatively high completion rate of material 3 and 4 for PEERs in Section A.

\begin{table*}[h!]
\centering
\begin{tabular}{lccc}
\hline\hline
& Coefficient & Standard error  & $p$ value\\
\hline
\textbf{Intercept}  &  -1.00  &   0.23  &     $< 0.001$  \\
\textbf{Section} & -1.16 & 0.27 &    $< 0.001$  \\
\textbf{(Reference = Section A)} &&& \\
\textbf{Gender} &   0.77  &  0.27    &    0.004     \\
\textbf{(Reference = Man)} &&&\\
\textbf{Race/ethnicity}   &   -0.78  &    0.39    &     0.045        \\
\textbf{(Reference  = WA)} &&&\\
Force concept inventory  &   -0.19  &      0.16   &   0.243        \\
Calculus concept inventory  &    0.24  &     0.16    &     0.131       \\
\hline\hline
\end{tabular}
\caption{Logistic regression model for math material 1 and 2 completion, where the completion variable is 1 if the student completes both materials. Variables with statistical significance are marked in bold.}
\label{logistic12}
\end{table*}

\begin{table*}[h!]
\centering
\begin{tabular}{lccc}
\hline\hline
& Coefficient & Standard error  & $p$ value\\
\hline
\textbf{Intercept}  &  -2.00  &   0.27  &     $< 0.001$  \\
\textbf{Section} & 1.06 & 0.28 &    $< 0.001$  \\
\textbf{(Reference = Section A)} &&& \\
Gender &  0.01  &  0.27    &    0.974     \\
(Reference = Man) &&&\\
Race/ethnicity   &   0.14 &    0.34    &     0.687        \\
(Reference  = WA) &&&\\
Force concept inventory  &  0.10  &      0.16   &   0.548       \\
Calculus concept inventory  &    -0.18  &     0.16    &     0.244      \\
\hline\hline
\end{tabular}
\caption{Logistic regression model for math material 3 and 4 completion, where the completion variable is 1 if the student completes both materials. Variables with statistical significance are marked in bold.}
\label{logistic34}
\end{table*}

\begin{table*}[h!]
\centering
\begin{tabular}{lccc}
\hline\hline
& Coefficient & Standard error  & $p$ value\\
\hline
\textbf{Intercept}  &      -1.24  &     0.37     &       0.001  \\
Gender &      0.01  &        0.08     &         0.895       \\
(Reference = Man) &&&\\
\textbf{Race/ethnicity}   &      -0.33  &        0.12    &         0.004        \\
\textbf{(Reference  = WA)} &&&\\
\textbf{Homework score}    &    1.38  &        0.38     &      $< 0.001$   \\
AI hints usage         &       -0.18  &        0.20     &      0.337     \\
Math supplement material 1 &       0.13  &        0.08      &     0.122   \\
\textbf{Force concept inventory}  &       0.42  &        0.05    &         $< 0.001$        \\
\textbf{Calculus concept inventory}    &       0.20  &        0.06      &         $< 0.001$       \\
\hline\hline
\end{tabular}
\caption{OLS regression model for Exam 1 score. Variables with statistical significance are marked in bold.}
\label{exam1ols}
\end{table*}

\begin{table*}[h!]
\centering
\begin{tabular}{lccc}
\hline\hline
& Coefficient & Standard error  & $p$ value\\
\hline
\textbf{Intercept}  &      -3.20  &       0.31     &     $<0.001$  \\
Gender &      0.00  &        0.08     &         0.967      \\
(Reference = Man) &&&\\
\textbf{Race/ethnicity}   &      -0.32  &      0.105    &      0.002        \\
\textbf{(Reference  = WA)} &&&\\
\textbf{Homework score}    &       3.34  &     0.32     &      $<0.001$   \\
\textbf{AI hints usage}     &       0.40  &        0.14     &      0.004     \\
Math supplement material 1 &       0.07  &     0.08      &    0.360      \\
Math supplement material 3 &      0.07  &      0.08     &     0.372      \\
\textbf{Force concept inventory}  &    0.29  &        0.05    &      $<0.001$        \\
\textbf{Calculus concept inventory}    &     0.15  &        0.05      &      0.003       \\
\hline\hline
\end{tabular}
\caption{OLS regression model for Exam 2 score. Variables with statistical significance are marked in bold.}
\label{exam2ols}
\end{table*}

\begin{table*}[h!]
\centering
\begin{tabular}{lccc}
\hline\hline
  & Coefficient & Standard error  & $p$ value\\
\hline
\textbf{Intercept}   &  -3.18  &   0.31    & $<0.001$   \\
Gender     &   0.01  &   0.08  &   0.875  \\
(Reference = Man) &&&\\
\textbf{Race/ethnicity}    &  -0.32  &   0.11   &  0.002  \\
\textbf{(Reference = WA)} &&&\\
\textbf{Homework score}    &  3.32  &   0.32   &   $<0.001$   \\
Math supplement material 1 &   0.07    &    0.08    &  0.353    \\
Math supplement material 3 &   0.07  &    0.08    &  0.399   \\
\textbf{Calculus concept inventory}     &  0.15   &  0.05     &  0.004   \\
\textit{\textbf{AI hints usage}}       &   0.33  &    0.13    &  0.008   \\
\textit{\textbf{Force concept inventory}}  &  0.33   &  0.06   &   $<0.001$  \\
\textit{AI hints usage : Force concept}  &   -0.31    &   0.18    &  0.094  \\
\hline\hline
\end{tabular}
\caption{Regression model on Exam 2 with interaction term. The last row corresponds to the interaction term. Statistically significant variables are marked in bold and the relevant interaction variables are marked in italics. The negative coefficient for the interaction term indicates that the positive effect of using AI hints is stronger for less prepared students.}
\label{exam2int}
\end{table*}

\end{document}